%% file: collage.tex
\documentclass{article}

\usepackage{microtype}
\usepackage{graphicx}
\usepackage{subcaption}
\usepackage{algorithm,algorithmic}
\usepackage{booktabs} % for professional tables
\usepackage{hyperref}

\usepackage[accepted]{icml2019}
\icmltitlerunning{Collage Inference}

\begin{document}
\twocolumn[
\icmltitle{Collage Inference: Achieving low tail latency during distributed image classification using coded redundancy models}




\begin{icmlauthorlist}
\icmlauthor{Krishna Giri Narra}{usc}
\icmlauthor{Zhifeng Lin}{usc}
\icmlauthor{Ganesh Ananthanarayanan}{msr}
\icmlauthor{Salman Avestimehr}{usc}
\icmlauthor{Murali Annavaram}{usc}
\end{icmlauthorlist}

\icmlaffiliation{usc}{EE Department, University Of Southern California, Los Angeles, California, USA}
\icmlaffiliation{msr}{Microsoft Research, Redmond, Washington, USA}

\icmlcorrespondingauthor{Krishna Giri Narra}{narra@usc.edu}

\icmlkeywords{Machine Learning, ICML}

\vskip 0.3in
]



\printAffiliationsAndNotice{}

\begin{abstract}
\input{0.abstract}
\end{abstract}

\section{Introduction}
\label{sec:introduction}
\input{1.introduction}

%\section{Motivation}
%\label{sec:background}
%\input{2.background}

\section{Collage Inference Technique}
\label{sec:collage_inference}
\input{3.collage_inference}

\section{Experimental Results}
\label{sec:evaluations}
\input{5.evaluations}

\section{Conclusion and Future Work}
\label{sec:conclusion}
\input{7.conclusion}

\bibliographystyle{ACM-Reference-Format}
\bibliography{collage.bib}
\end{document}

%% file: 0.abstract.tex
Reducing the latency variance in machine learning inference is a key requirement in many applications. Variance is harder to control in a cloud deployment in the presence of stragglers. In spite of this challenge, inference is increasingly being done in the cloud, due to the advent of affordable machine learning as a service (MLaaS) platforms. Existing approaches to reduce variance rely on replication which is expensive and partially negates the affordability of MLaaS.  In this work, we argue that MLaaS platforms also provide unique opportunities to cut the cost of redundancy. In MLaaS platforms, multiple inference requests are concurrently received by a load balancer which can then create a more cost-efficient redundancy coding across a larger collection of images.
We propose a novel convolutional neural network model, Collage-CNN, to provide a low-cost redundancy framework. A Collage-CNN model takes a collage formed by combining multiple images and performs multi-image classification in one shot, albeit at slightly lower accuracy. We then augment a collection of traditional single image classifiers with a single Collage-CNN classifier which acts as a low-cost redundant backup. Collage-CNN then provides backup classification results if a single image classification straggles. Deploying the Collage-CNN models in the cloud, we demonstrate that the 99th percentile tail latency of inference can be reduced by 1.47X compared to replication based approaches while providing high accuracy. Also, variation in inference latency can be reduced by 9X with a slight increase in average inference latency.

%Machine learning as a service (MLaaS) platforms are increasingly used to perform training and inference of ML models. The inference is performed by deploying the trained models on the MLaas platforms to serve queries from users or applications. To achieve scalability of the inference service, incoming queries are distributed across multiple replicas of the ML model. Building such a scalable MLaaS that has low tail latency is challenging due to the incidence of straggler nodes. In this paper, we propose a novel coded redundancy model to deal with stragglers in distributed image classification. We propose a novel convolutional neural network model, Collage-CNN, to provide efficient resilience in the presence of stragglers. A Collage-CNN model takes collage images formed by combining multiple images as its input and performs multi-image classification in one shot. We generate custom training collages using images from standard image classification datasets and train the Collage-CNN model to achieve high multi-image classification accuracy. Deploying the Collage-CNN models in the cloud, we demonstrate that the 99th percentile tail latency of inference can be reduced by 1.47X compared to replication based approaches while providing high accuracy. Also, variation in inference latency can be reduced by 9X with a slight increase in average inference latency.

%% file: 1.introduction.tex
Low latency and low variance machine learning inference is critical in many control systems applications, such as robotics.  Machine learning as a service (MLaaS) platforms are attractive for scaling inference traffic. The inference is performed by deploying the trained models on the MLaaS platforms. To achieve scalability of the inference service, incoming queries are distributed across multiple replicas of the ML model. As the inference demands grow an enterprise can simply increase the cloud instances to meet the demand.  However, virtualized services are prone to straggler problems, which lead to the high variability and long tail in the inference latency. Straggler incidence is more acute in cloud-based deployments because of the widespread sharing of compute, memory and network resources~\cite{tail}.

\begin{figure}
    \centering
    \includegraphics[width=\linewidth, height=4cm]{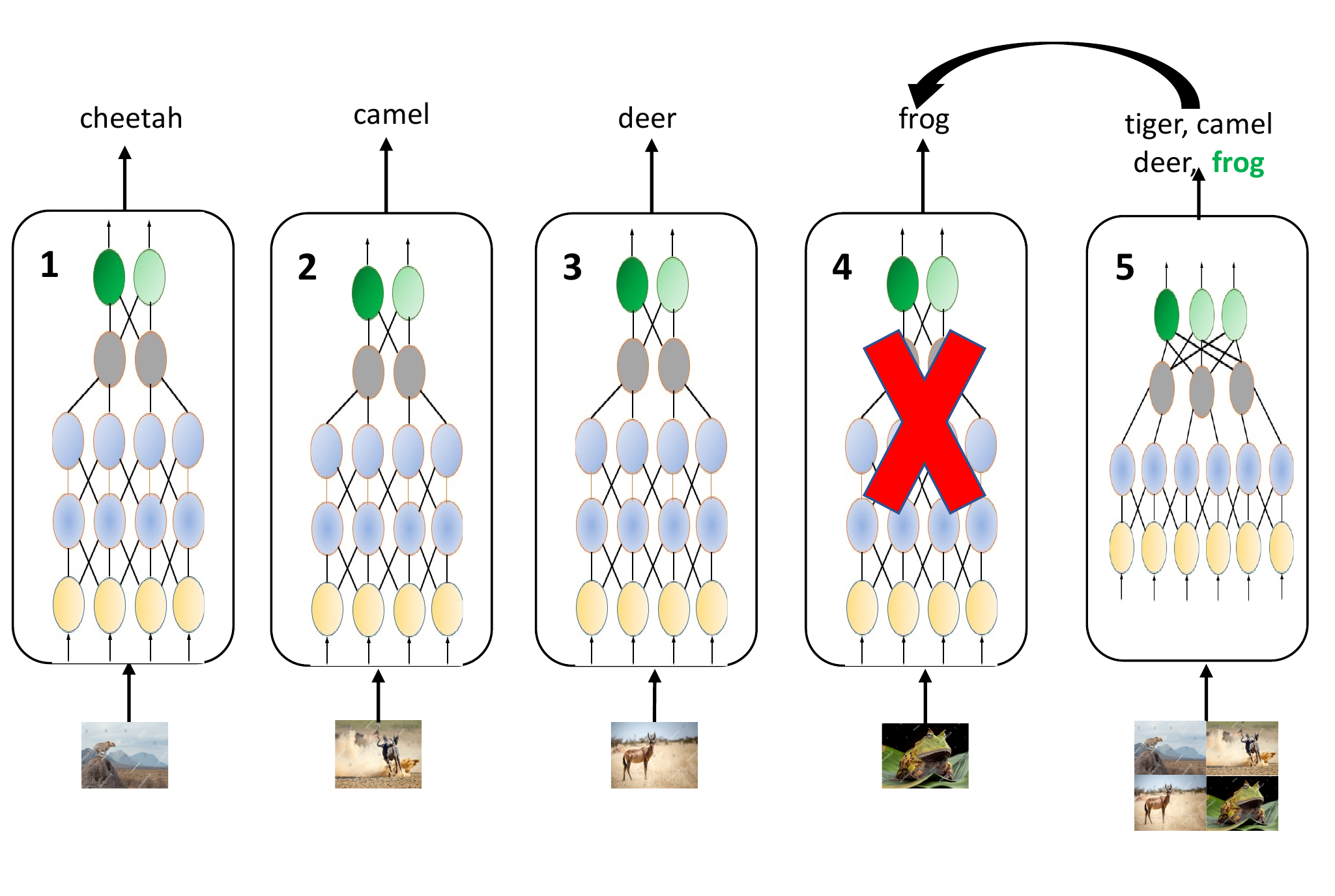}
    \caption{Collage Inference illustration}
    \label{fig:introduction_collage}
\end{figure}

Existing techniques to lower tail latency can be broadly classified into two categories: replication (e.g., \cite{tail, ananthanarayanan2010reining, wang}), coded computing (e.g.,\cite{speedUpML,unifiedCoding,dutta2016short,yu2017polynomial}). In replication based techniques, additional resources are used to add redundancy during execution: either a task is replicated at its launch, or a task is replicated on detection of a straggler node. Replicating every request pro-actively as a straggler mitigation strategy could lead to a significant increase in resource costs. Replicating a request reactively on the detection of a straggler can increase latency. While MLaaS platforms are more prone to stragglers, in this work we argue that they are also more amenable to low cost redundancy schemes. MLaaS platforms deploy a front-end load balancer that receives requests from multiple users and submits them to the back-end cloud instances. In this setting, the load balancer has the unique advantage of treating multiple requests as a single collective and create a more cost effective redundancy strategy. 

%Coded computing techniques add redundancy in a coded form at the launch of tasks and have proven useful for linear computing tasks. 
%Coded computing has been recently applied to mitigate stragglers in image classification \cite{learning_a_code}, but the proposed technique suffers from a significant drop in accuracy.

We propose the Collage Inference technique as a cost effective redundancy strategy to deal with variance in inference latency. Collage Inference uses a unique convolutional neural network (CNN) based coded redundancy model, referred to as a Collage-CNN, that can perform multiple predictions in one shot, albeit at a slight reduction in prediction accuracy. Collage-CNN is like a parity model where the input encoding is the collection of images that are spatially arranged into a collage. Its output is decoded to get the missing predictions of images that are assigned to straggler nodes. This coded redundancy model is run concurrently as a single backup service for a collection of individual image inference models. An individual image inference model is referred to as an S-CNN. Figure \ref{fig:introduction_collage} shows a service comprising of four S-CNN models and one Collage-CNN model. When prediction from model 4 is missing, the corresponding prediction from Collage-CNN is used in its place.

%% file: 3.collage_inference.tex
\begin{figure}
    \centering
    \includegraphics[width=\linewidth]{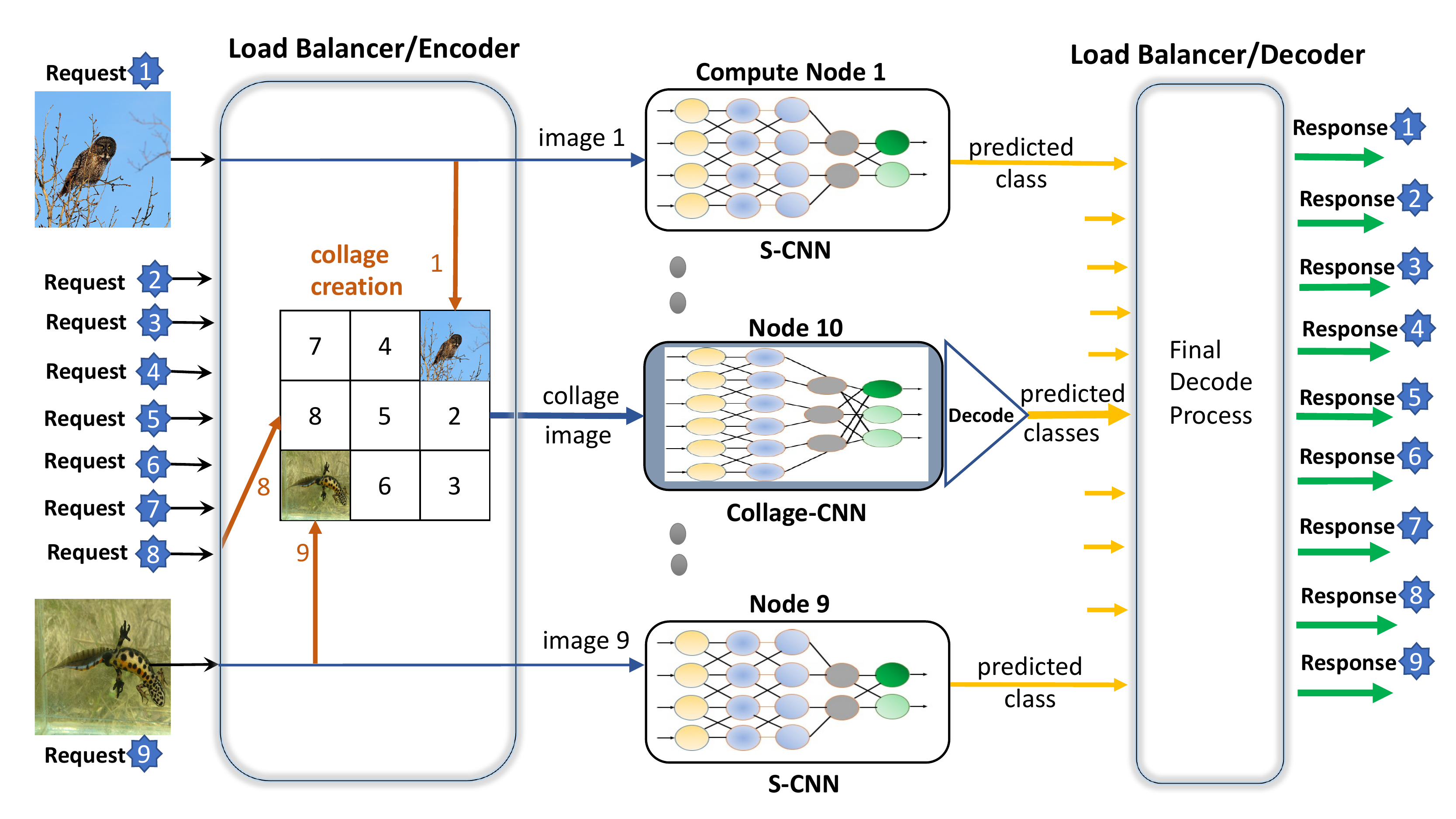}
    \caption{Collage Inference Technique}
    \label{fig:collage_inference_algorithm}
\end{figure}

The Collage-CNN model is a novel multi-object classification model. The critical insight behind collage inference is that the spatial information within an input image is critical, for CNNs to achieve high accuracy, and it should be maintained. So, we use the collage image composed of all the images as the encoding. The Collage-CNN model takes a collage encoded from the images $[Image_1,.., Image_i, .., Image_N]$, where each image is input to one of the $N$ single image classifiers. The Collage-CNN provides the predictions for all the objects in the collage along with the locations of each object in the collage. The predicted locations are in the form of rectangular bounding boxes. By smartly encoding the individual images into a collage and using location information from the Collage-CNN predictions, the collage inference technique can replace the missing predictions from any straggler nodes.

The encoding of individual images into a single collage image happens as follows. Let a Collage-CNN be providing backup for $N$ S-CNN model replicas that are each running on a compute node. To encode the $N$ images into a collage we first create a square grid consisting of $[\sqrt{N}, \sqrt{N}]$ boxes. Each image that is assigned to an S-CNN model running on compute node $i$ is placed in  a predefined square box within the collage. Specifically, in the collage, each compute node $i$ is assigned the box location $i$. This encoding information is used while decoding outputs of the Collage-CNN. From the outputs of the Collage-CNN, class prediction corresponding to each bounding box $i$ is extracted, and this prediction corresponds to the node $i$. Our Collage-CNN model takes collage with 416x416 resolution as input. As the size of $N$ grows, more images must be packed into the collage, which reduces the resolution of each image. It can lower the accuracy of predictions.

Figure \ref{fig:collage_inference_algorithm} shows the collage inference technique for ten nodes with one of the nodes providing redundancy for the remaining $N=9$ nodes. Each of the nine nodes running S-CNN model takes an individual image as input. The $10^{th}$ node takes the collage image as input. Inside the load balancer, each of the nine input images is lowered in resolution and inserted into a specific location to form the collage image. The input image to node $i$ goes into location $i$ in the collage image. 
%This placement order is useful later on to map different objects predicted by the Collage-CNN to corresponding input images.
This collage image is provided as input to node 10. The predictions from the Collage-CNN are processed using the collage decoding algorithm. The output predictions from all the ten nodes go to the final decode process in the load balancer. This decode process uses the predictions from the Collage-CNN model to fill in any missing predictions from the nine nodes and return the final prediction responses to the user.

\begin{figure}
    \centering
    \begin{subfigure}{0.3\linewidth}
    \includegraphics[width=\linewidth]{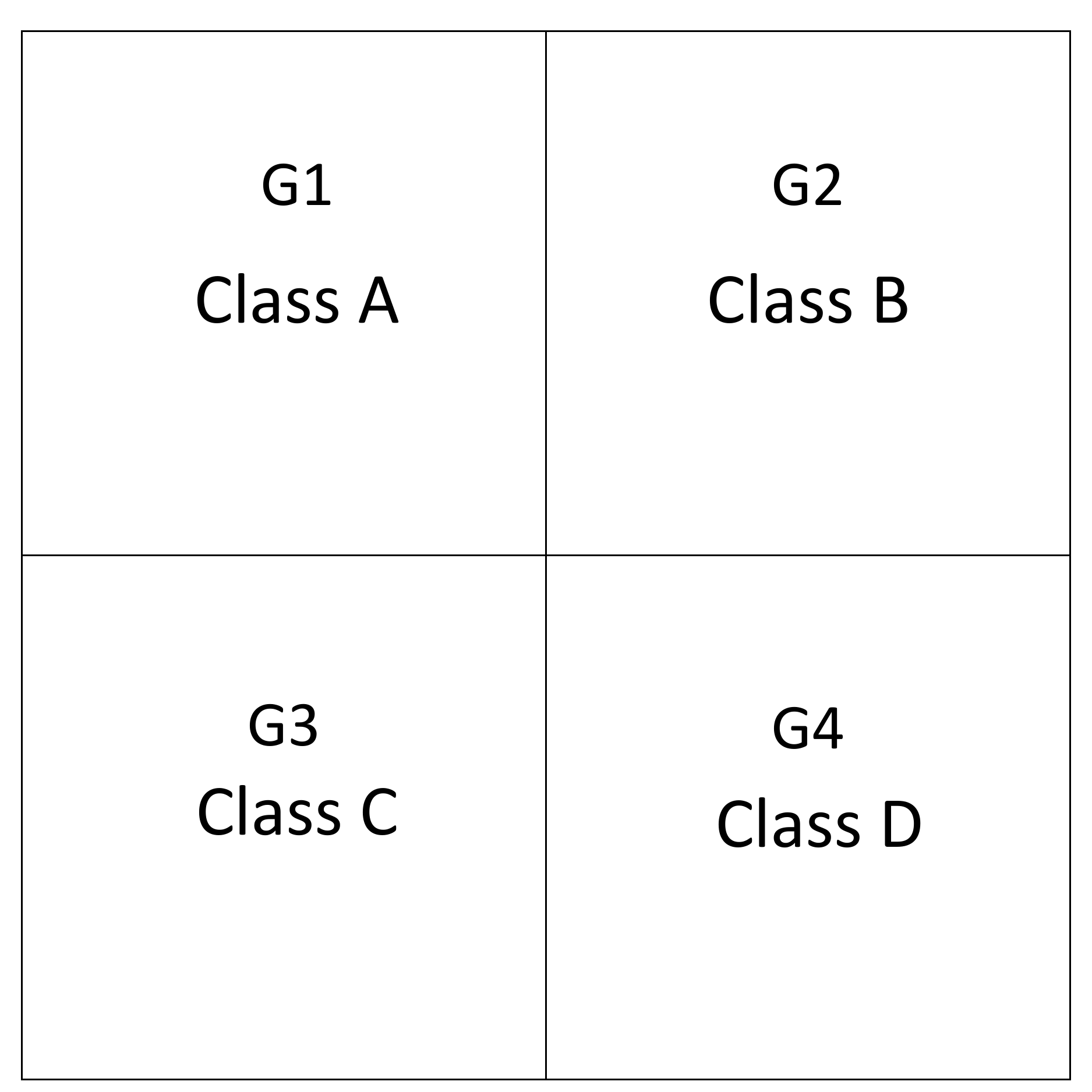}
    \caption{Ground Truth}
    \end{subfigure}
    \begin{subfigure}{0.3\linewidth}
    \includegraphics[width=\linewidth]{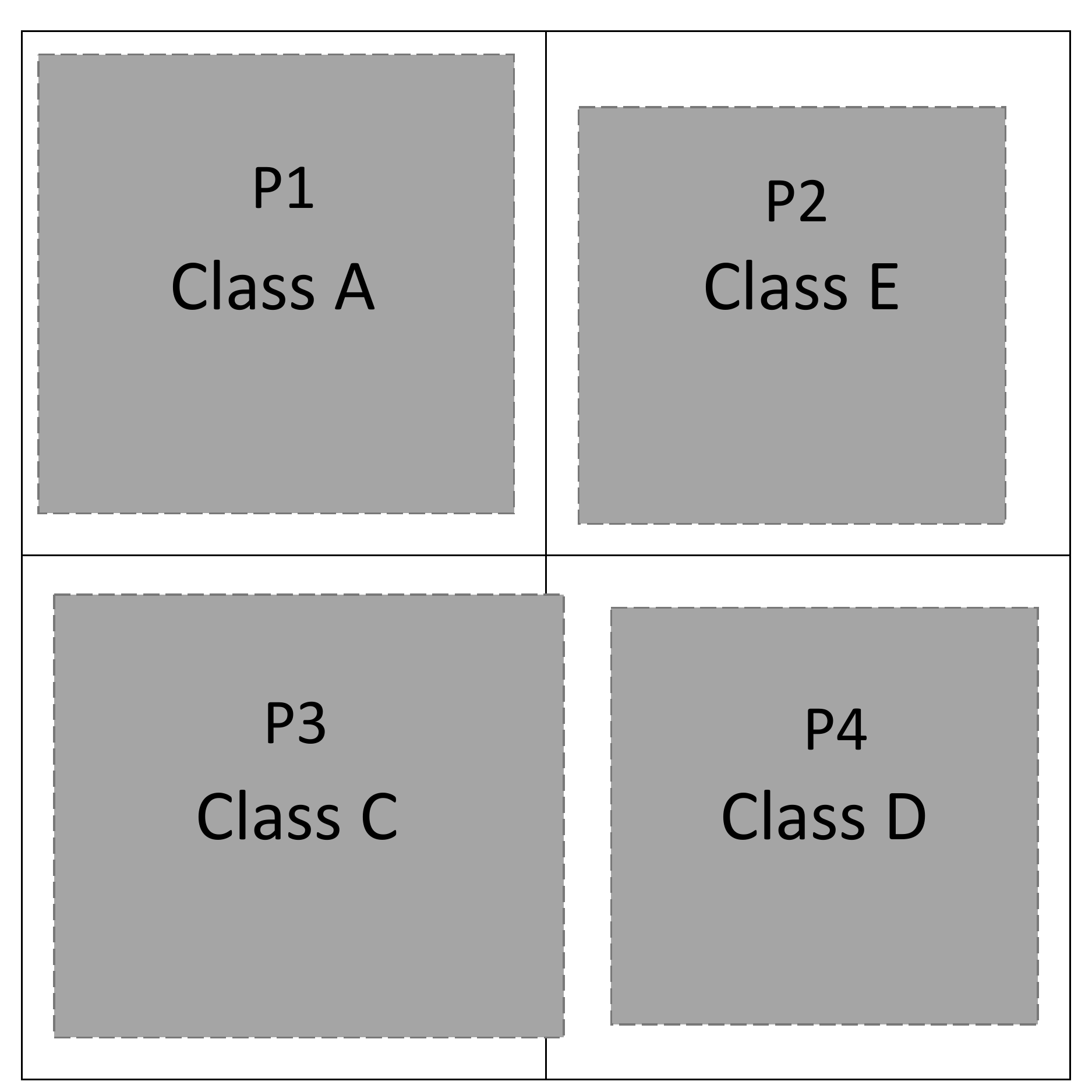}
    \caption{Scenario 1}
    \end{subfigure}
    \begin{subfigure}{0.3\linewidth}
    \includegraphics[width=\linewidth]{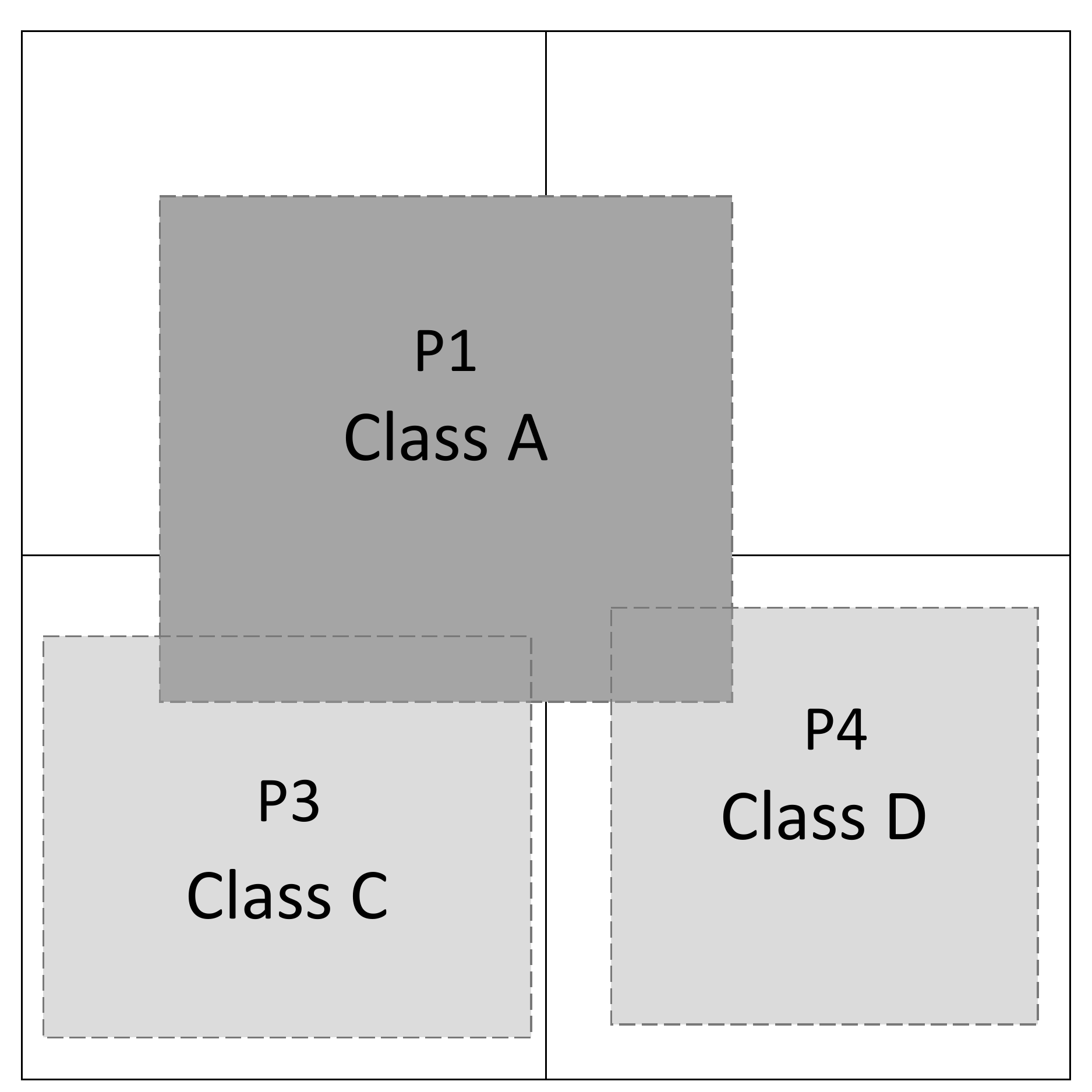}
    \caption{Scenario 2}
    \end{subfigure}
    \caption{Few Collage Prediction scenarios}
    \label{fig:collage_output_scenarios}
\end{figure}

The collage decoding algorithm extracts the best possible class predictions for the $N$ images from all the Collage-CNN predictions. The decoding algorithm calculates the Jaccard similarity coefficient of each predicted bounding box with each of the $N$ ground truth bounding boxes that are used in creating the collages. Let area of ground truth bounding box be $A_{gt}$, area of predicted bounding box be $A_{pred}$ and area of intersection between both the boxes be $A_i$. Then Jaccard similarity coefficient can be computed using the formula: $\frac{A_i}{A_{gt} + A_{pred} - A_i}$. The ground truth bounding box with the largest similarity coefficient is assigned the class label of the predicted bounding box. As a result, the image present in this ground truth bounding box is predicted as having an object belonging to this class label. This is repeated for all the object predictions. To illustrate the algorithm, consider example scenarios shown in figure \ref{fig:collage_output_scenarios}. The ground truth input collage is a 2x2 collage that is formed from four images. It has four bounding boxes G1, G2, G3, and G4 which contain objects belonging to classes A, B, C, and D respectively. In scenario 1, the collage model predicts four bounding boxes P1, P2, P3 and P4. In this scenario: P1 would have largest similarity value with G1, P2 with G2, P3 with G3 and P4 with G4. So, the decoding algorithm predicts class A in G1, class E in G2, class C in G3, class D in G4. In scenario 2, three bounding boxes are predicted by the model. Predicted box P1 is spread over G1, G2, G3 and G4. Jaccard similarity value of P1 with box G1 is: $\frac{1}{3}$, G2 is: $\frac{1}{7}$, G3 is: $\frac{1}{7}$ and G4 is: $\frac{1}{17}$. So, the algorithm predicts class A in G1, empty prediction in G2, class C in G3, class D in G4.

%% file: 5.evaluations.tex
We trained and measured the top-1 accuracy of Collage-CNN and S-CNN models using images from 100 classes of the Imagenet-1k (ILSVRC 2012) dataset.  Resnet-34 is used as the S-CNN model, and its accuracy is 80.72\%. Accuracy of Collage-CNN is 76.9\% when there are nine images per collage. Hence, Collage-CNN essentially tradesoff a small accuracy degradation to improve the cost of redundancy through collage coding.

We implemented an online image classification system and deployed it on the Digital Ocean cloud. This system is similar to the one described in figure \ref{fig:collage_inference_algorithm} where a load balancer receives requests from multiple clients concurrently. The load balancer is responsible for creating an appropriate collage image with the incoming images. We performed experiments with nine S-CNN compute nodes and one Collage-CNN compute node. Validation images from Imagenet dataset are used to generate inference requests. Two baselines are used for comparison. First is the no replication baseline, where no straggling request is replicated. Second is the replication baseline, where straggling requests are replicated based on a fixed timeout. We measured the end-to-end latency for each request from the time it is sent to the time predictions for it are received. For requests to Collage-CNN model, the end-to-end latency also includes time spent in creating the collage image.

\begin{figure}
    \centering
    \includegraphics[width=\linewidth]{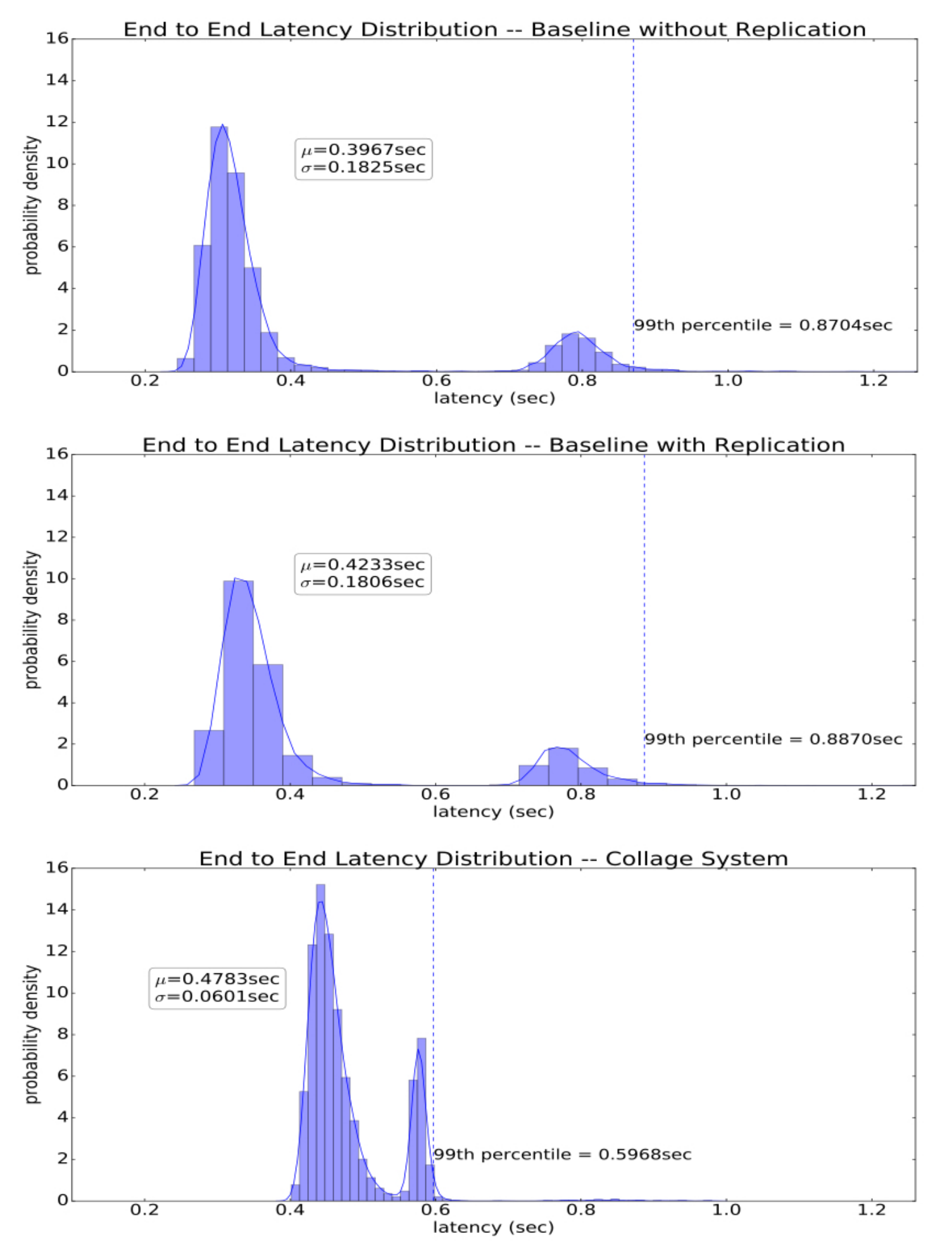}
    \caption{Inference latency comparison}
    \label{fig:3x3_all}
\end{figure}

The end to end latency distribution observed when the image classification system consists of nine S-CNN models with no request replication is shown in the top sub-figure of figure \ref{fig:3x3_all}. The middle sub-figure corresponds to the system consisting of nine S-CNN models with request replication. The bottom sub-figure corresponds to system consisting of nine S-CNN models and one Collage-CNN model. The x-axis shows the latency in seconds. The histograms along y-axis are the probability density values for the latency distribution. The blue curve line shows the estimated probability density function. Collage inference has a slightly higher mean latency due to the collage creation time. Using Collage-CNN model reduces the standard deviation in latency by 3X and variance by 9X. The 99-th percentile latency of Collage inference is 1.47X lower than both No replication and Replication methods. When the Collage-CNN predictions are used by the final decoder to fill in the missing predictions, the accuracy of these predictions is 87.86\%. It is significantly better than the top-1 accuracy (76.9\%) because, when using Collage-CNN, only a subset of its predictions corresponding to the straggler nodes need to be accurate.

%% file: 7.conclusion.tex
In this paper we described a novel coded redundancy model and demonstrated that it reduces inference tail latency. Future work includes improving the Collage-CNN model and reducing the overhead of creating the collage image.

%% file: collage.bbl

\begin{thebibliography}{00}

%%% the Web address.

\ifx \showCODEN    \undefined \def \showCODEN     #1{\unskip}     \fi
\ifx \showDOI      \undefined \def \showDOI       #1{#1}\fi
\ifx \showISBNx    \undefined \def \showISBNx     #1{\unskip}     \fi
\ifx \showISBNxiii \undefined \def \showISBNxiii  #1{\unskip}     \fi
\ifx \showISSN     \undefined \def \showISSN      #1{\unskip}     \fi
\ifx \showLCCN     \undefined \def \showLCCN      #1{\unskip}     \fi
\ifx \shownote     \undefined \def \shownote      #1{#1}          \fi
\ifx \showarticletitle \undefined \def \showarticletitle #1{#1}   \fi
\ifx \showURL      \undefined \def \showURL       {\relax}        \fi
% invisible to TeX
\providecommand\bibfield[2]{#2}
\providecommand\bibinfo[2]{#2}
\providecommand\natexlab[1]{#1}
\providecommand\showeprint[2][]{arXiv:#2}

\bibitem[\protect\citeauthoryear{Ananthanarayanan, Kandula, Greenberg, Stoica,
  Lu, Saha, and Harris}{Ananthanarayanan et~al\mbox{.}}{2010}]%
        {ananthanarayanan2010reining}
\bibfield{author}{\bibinfo{person}{Ganesh Ananthanarayanan},
  \bibinfo{person}{Srikanth Kandula}, \bibinfo{person}{Albert~G Greenberg},
  \bibinfo{person}{Ion Stoica}, \bibinfo{person}{Yi Lu}, \bibinfo{person}{Bikas
  Saha}, {and} \bibinfo{person}{Edward Harris}.}
  \bibinfo{year}{2010}\natexlab{}.
\newblock \showarticletitle{Reining in the Outliers in Map-Reduce Clusters
  using Mantri.}. In \bibinfo{booktitle}{{\em OSDI}},
  Vol.~\bibinfo{volume}{10}. \bibinfo{pages}{24}.
\newblock


\bibitem[\protect\citeauthoryear{Dean and Barroso}{Dean and Barroso}{2013}]%
        {tail}
\bibfield{author}{\bibinfo{person}{Jeffrey Dean} {and}
  \bibinfo{person}{Luiz~André Barroso}.} \bibinfo{year}{2013}\natexlab{}.
\newblock \showarticletitle{The Tail at Scale}.
\newblock \bibinfo{journal}{{\it Commun. ACM}}  \bibinfo{volume}{56}
  (\bibinfo{year}{2013}), \bibinfo{pages}{74--80}.
\newblock
\showURL{%
\url{http://cacm.acm.org/magazines/2013/2/160173-the-tail-at-scale/fulltext}}


\bibitem[\protect\citeauthoryear{Dutta, Cadambe, and Grover}{Dutta
  et~al\mbox{.}}{2016}]%
        {dutta2016short}
\bibfield{author}{\bibinfo{person}{Sanghamitra Dutta}, \bibinfo{person}{Viveck
  Cadambe}, {and} \bibinfo{person}{Pulkit Grover}.}
  \bibinfo{year}{2016}\natexlab{}.
\newblock \showarticletitle{Short-Dot: Computing Large Linear Transforms
  Distributedly Using Coded Short Dot Products}. In \bibinfo{booktitle}{{\em
  Advances In Neural Information Processing Systems}}.
  \bibinfo{pages}{2092--2100}.
\newblock


\bibitem[\protect\citeauthoryear{Lee, Lam, Pedarsani, Papailiopoulos, and
  Ramchandran}{Lee et~al\mbox{.}}{2016}]%
        {speedUpML}
\bibfield{author}{\bibinfo{person}{Kangwook Lee}, \bibinfo{person}{Maximilian
  Lam}, \bibinfo{person}{Ramtin Pedarsani}, \bibinfo{person}{Dimitris
  Papailiopoulos}, {and} \bibinfo{person}{Kannan Ramchandran}.}
  \bibinfo{year}{2016}\natexlab{}.
\newblock \showarticletitle{Speeding up distributed machine learning using
  codes}. In \bibinfo{booktitle}{{\em 2016 IEEE International Symposium on
  Information Theory (ISIT)}}. \bibinfo{pages}{1143--1147}.
\newblock
\showDOI{%
\url{https://doi.org/10.1109/ISIT.2016.7541478}}


\bibitem[\protect\citeauthoryear{{Li}, {Maddah-Ali}, and {Avestimehr}}{{Li}
  et~al\mbox{.}}{2016}]%
        {unifiedCoding}
\bibfield{author}{\bibinfo{person}{S. {Li}}, \bibinfo{person}{M.~A.
  {Maddah-Ali}}, {and} \bibinfo{person}{A.~S. {Avestimehr}}.}
  \bibinfo{year}{2016}\natexlab{}.
\newblock \showarticletitle{A Unified Coding Framework for Distributed
  Computing with Straggling Servers}. In \bibinfo{booktitle}{{\em 2016 IEEE
  Globecom Workshops (GC Wkshps)}}. \bibinfo{pages}{1--6}.
\newblock
\showDOI{%
\url{https://doi.org/10.1109/GLOCOMW.2016.7848828}}


\bibitem[\protect\citeauthoryear{Wang, Joshi, and Wornell}{Wang
  et~al\mbox{.}}{2014}]%
        {wang}
\bibfield{author}{\bibinfo{person}{Da Wang}, \bibinfo{person}{Gauri Joshi},
  {and} \bibinfo{person}{Gregory Wornell}.} \bibinfo{year}{2014}\natexlab{}.
\newblock \showarticletitle{Efficient Task Replication for Fast Response Times
  in Parallel Computation}.
\newblock \bibinfo{journal}{{\em SIGMETRICS Perform. Eval. Rev.\/}}
  \bibinfo{volume}{42}, \bibinfo{number}{1} (\bibinfo{date}{June}
  \bibinfo{year}{2014}), \bibinfo{pages}{599--600}.
\newblock
\showISSN{0163-5999}
\showDOI{%
\url{https://doi.org/10.1145/2637364.2592042}}


\bibitem[\protect\citeauthoryear{Yu, Maddah-Ali, and Avestimehr}{Yu
  et~al\mbox{.}}{2017}]%
        {yu2017polynomial}
\bibfield{author}{\bibinfo{person}{Qian Yu}, \bibinfo{person}{Mohammad
  Maddah-Ali}, {and} \bibinfo{person}{Salman Avestimehr}.}
  \bibinfo{year}{2017}\natexlab{}.
\newblock \showarticletitle{Polynomial codes: an optimal design for
  high-dimensional coded matrix multiplication}. In \bibinfo{booktitle}{{\em
  Advances in Neural Information Processing Systems}}.
  \bibinfo{pages}{4403--4413}.
\newblock


\end{thebibliography}
